# Local formation of nitrogen-vacancy centers in diamond by swift heavy ions

J. Schwartz[1, 3], S. Aloni[2], D. F. Ogletree[2], M. Tomut[4], M. Bender[4], D. Severin[4], C. Trautmann[4, 5], I. W. Rangelow[3], and T. Schenkel[1]

[1]Accelerator and Fusion Research Division, [2]The Molecular Foundry, Lawrence Berkeley National Laboratory, Berkeley CA 94720, USA

[3]Ilmenau University of Technology, Department of Microelectronics and Nanoelectric Systems, 98684 Ilmenau, Germany

[4]GSI Helmholtz Center for Heavy Ion research, 64291 Darmstadt, Germany

[5]Technische Universität Darmstadt, 64287 Darmstadt, Germany

We exposed nitrogen-implanted diamonds to beams of swift heavy ions (~1 GeV, ~4 MeV/u) and find that these irradiations lead directly to the formation of nitrogen vacancy (NV) centers, without thermal annealing. We compare the photoluminescence intensities of swift heavy ion activated NV$^-$ centers to those formed by irradiation with low-energy electrons and by thermal annealing. NV$^-$ yields from irradiations with swift heavy ions are 0.1 of yields from low energy electrons and 0.02 of yields from thermal annealing. We discuss possible mechanisms of NV center formation by swift heavy ions such as electronic excitations and thermal spikes. While forming NV centers with low efficiency, swift heavy ions could enable the formation of three dimensional NV$^-$ assemblies over relatively large distances of tens of micrometers. Further, our results show that NV center formation is a local probe of (partial) lattice damage relaxation induced by electronic excitations from swift heavy ions in diamond.

## 1. Introduction

Negatively charged nitrogen vacancy centers (NV$^-$) in diamond are promising candidates for quantum information processing and for use as highly sensitive probes, e.g. in magnetometry [1-4]. In diamonds that were doped with nitrogen during the growth process NV centers can be formed by creating vacancies using irradiations with energetic photons, neutrons, electrons or ions followed by thermal annealing above 600° C. Here, vacancies become mobile and substitutional nitrogen atoms ($N_s$ or P1 centers) can capture single vacancies, forming NV centers. In pure diamonds, nitrogen can be added by implantation of nitrogen ions [5-8]. Ion implantation offers control over the local nitrogen and NV center concentrations with high spatial resolution [5, 9]. When implanting a diamond with nitrogen ions, e. g. of keV to MeV energies, ions transfer kinetic energy to carbon atoms in elastic collisions, leading to collision cascades that generate vacancies and carbon interstitials. Implanted nitrogen ions come to rest mostly on interstitial lattice sites, though replacement collisions are also possible. The formation of NV centers in nitrogen implanted diamond has been assumed to follow a simple two-step process. During annealing above 600°C, implanted nitrogen atoms can first be incorporated on substitutional lattice sites, forming P1 centers [10]. Vacancies can diffuse through the diamond lattice during thermal annealing and the second step in NV formation is the trapping of a vacancy at the site of a substitutional nitrogen atom [11, 12]. The charge state of the resulting NV centers is affected by the local nitrogen concentration, the defect environment and by surface conditions [11, 13]. Only NV$^0$ and NV$^-$ centers are observed in photoluminescence (PL) or cathodoluminescence (CL) measurements. Reported efficiencies for the conversion of implanted nitrogen atoms to NV centers depend on the nitrogen implant energy and processing conditions (including implantation with elevated target temperatures [8]) and range from below 1% to about 25% for low energy nitrogen implants (<100 keV) [8, 14, 15].

Hydrogen, present from growth or introduced into diamonds during annealing can interact with NV centers and quench their photoluminescence [16, 17].

Recent simulations of NV center formation in diamonds with high concentrations of substitutional nitrogen suggest that NV centers form during exposure to vacancy-producing irradiation and not during thermal annealing. The reason for this is the apparently higher likelihood for di-vacancy vs. NV formation during annealing [11].

In previous work [18], we had reported that NV centers also form in nitrogen-implanted, electronic grade diamond during exposure to low-energy electrons (1 – 30 keV) without any thermal annealing. This result showed that the commonly accepted two-step model of NV formation in electronic grade, nitrogen-implanted diamond is incomplete. We had hypothesized that defect complexes comprised of interstitial nitrogen atoms near several vacancies are present after nitrogen implantation and that electronic excitations from keV electron bombardment can induce reconstruction of this precursor into stable NV centers.

Here, we show that electronic excitations induced by swift heavy ions (SHI) also lead to the local formation of NV centers without thermal annealing. SHI such as gold or uranium ions with kinetic energies of ~1 GeV, near the Bragg peak in electronic energy loss, deposit kinetic energy mostly through inelastic collisions with target electrons at a rate of about 50 keV/nm in diamond. Energetic electrons quickly thermalize in cascades, forming thousands of electron-hole pairs per SHI. Lattice heating through electron – phonon coupling can lead to thermal spikes effects in the material surrounding the ion trajectory. We report relative yields of NV$^-$ centers formed by SHI compared to exposure to low energy electrons and compared to thermal annealing and discuss possible mechanisms of NV$^-$ formation by SHI. Insights into the interplay of electronic excitations and thermal annealing might lead to improved processing strategies for reliable formation of NV$^-$ centers with high placement resolution, spectral stability and favorable spin properties.

## 2. Experimental Methods

After cleaning in a mix of sulfuric, nitric and perchloric acid (1:1:1) we implanted two electronic grade diamond samples (<5 ppb nitrogen content, size of 4 mm$^2$, from Element 6) with $^{14}$N$^+$ ions. The implant spots were 1 mm in diameter and we aligned them so that they did not overlap. Each diamond was implanted with 10 and 30 keV nitrogen ions and for both implant energies two fluences, $10^{12}$ and $10^{13}$ N cm$^{-2}$ were applied for a total of four implant spots on each diamond. From earlier depth profile measurements we know the implantation depths for these implant conditions to be less than 130 nm, including channeling tails [15]. Irradiations with SHI were performed at the GSI Helmholtz Center using 1.14-GeV U ions of fluence 5×10$^{11}$ cm$^{-2}$ and 1.16-GeV Au ions (10$^{11}$ cm$^{-2}$). The projected range of these ions in diamond is about 30 μm, far beyond the 100 nm thick nitrogen-implanted layer [19]. The higher fluence, uranium-implanted sample showed a visible greenish discoloration presumably from vacancy related color centers, such as GR1 centers, formed by displacement damage preferentially at the end of the ion range. We focus here on NV$^-$ and did not track details of GR1 center formation and annealing [20].

During SHI irradiations we masked the samples with a honeycomb patterned metal grid with a thickness much larger than the range of the SHI. In the following, we concentrate our analysis on results from the uranium irradiated sample because signal levels from the sample irradiated with the lower fluence of gold ions were low.

In order to compare results from SHI irradiation to our previous results from low-energy electron bombardment, we exposed three different sample areas (pristine diamond, nitrogen implanted, nitrogen implanted + SHI irradiated) also to a high fluence of 10 keV electrons in a scanning electron microscope (SEM). The applied electron fluence was 90 C/cm$^2$ corresponding to saturation of the observed NV$^-$ yield reported earlier for diamonds that had been implanted with the same fluences of nitrogen ions [18]. We show a schematic of the implant conditions in Figure 1. The masked SHI irradiations resulted in sharp contrast in hyperspectral imaging between the neighboring irradiated and masked regions as seen in Figure 2.

For optical characterization we performed hyperspectral imaging at room temperature using a grating spectrometer Raman Microscope (Aramis by Horiba) with 0.8 NA air objective and a laser of wavelength 532 nm. All different exposure combinations were analyzed and compared by integrating the peak area around the NV$^-$ zero-phonon line (ZPL) before and after annealing at 850$^{\circ}$C. Figure 2 shows a hyperspectral image obtained by integration over the wavelength range of 635-642 nm around the ZPL of the NV$^-$ centers. The large circular zone (light red, indicated by the white line) corresponds to the spot of the nitrogen implantation. Regions covered with the honeycomb shaped metal mask during SHI irradiation appear as dark lines. The dark area (1) outside the nitrogen implant spot thus corresponds to pristine diamond that was neither implanted with nitrogen nor exposed to SHI. Area (2) outside the implantation spot was exposed only to 1.14 GeV uranium ions ($5\times10^{11}$ cm$^{-2}$) and shows a PL signal from single vacancies (GR1 center) caused by lattice displacements from the SHI impacts, but no NV$^-$ signal. Region (3) was implanted with nitrogen as well as exposed to SHI and shows NV$^-$ centers with a relatively low intensity. Areas that were implanted with nitrogen but then masked from SHI impacts (4) show NV$^-$ only after irradiation with low-energy electrons (as in [18]). We also irradiated N-implanted and SHI-irradiated areas in micron scale spots with a 10-keV electron beam (5) to explore possible additive effects in NV formation from consecutive SHI and electron exposures.

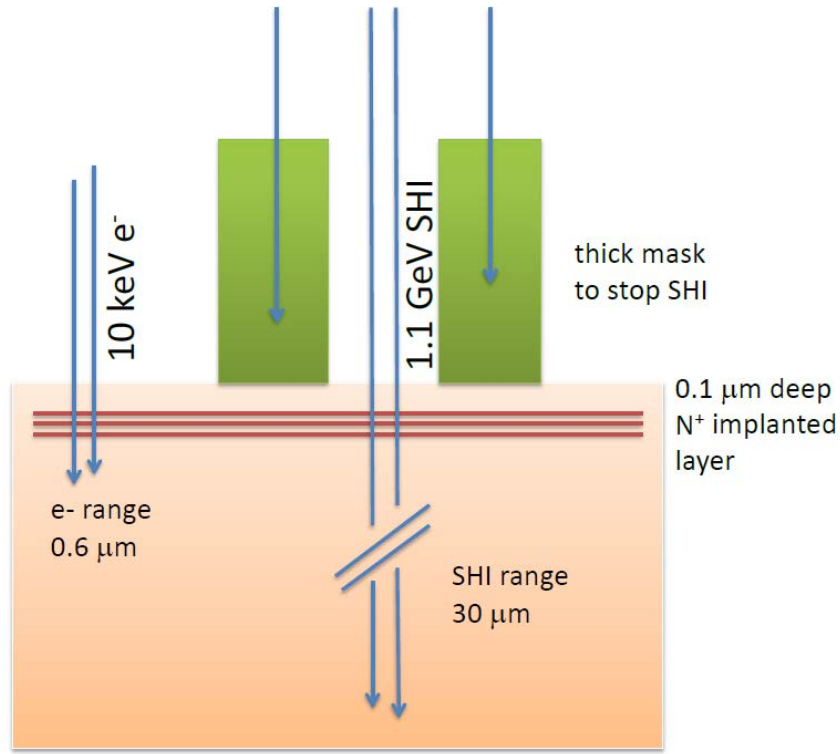

Figure 1: Schematic of the irradiation conditions.

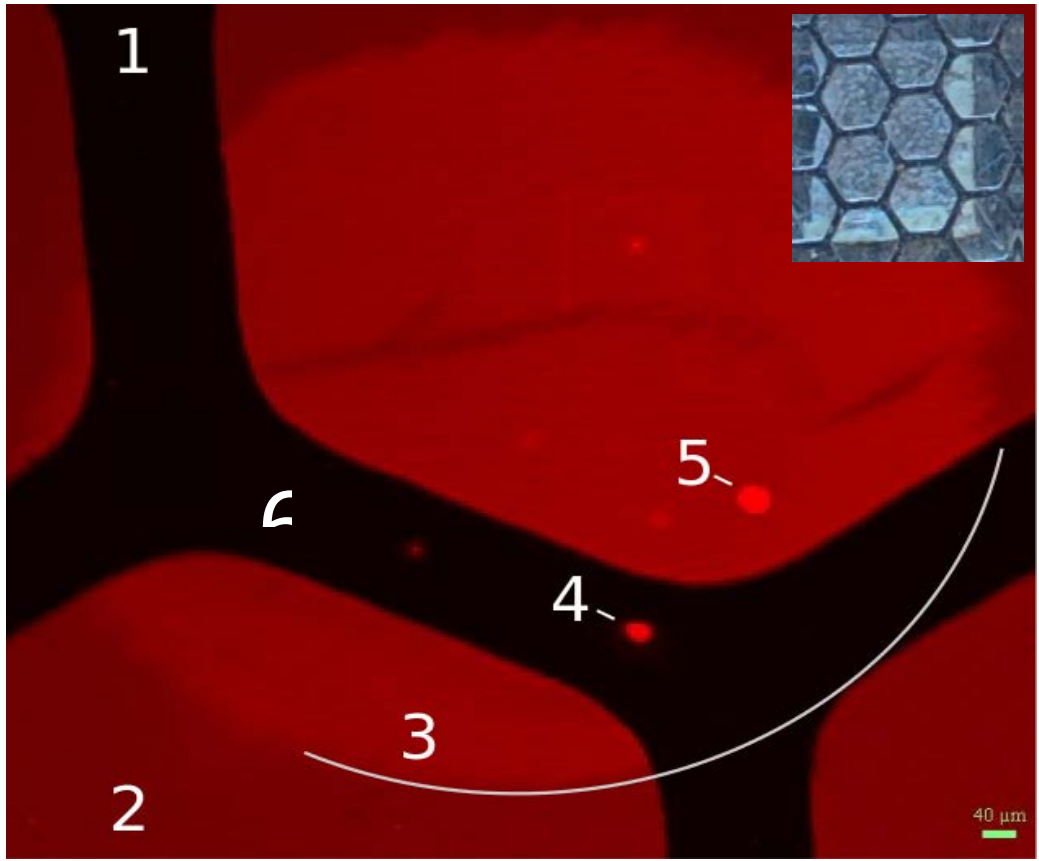

Figure 2: Hyperspectral image of the PL intensity from NV⁻ centers (635-642 nm) before thermal annealing. The central region was implanted with nitrogen ions (10 keV, $10^{13}$ cm$^{-2}$, indicated by the white line). During subsequent SHI irradiation (1.14 GeV U, $5\times10^{11}$ cm$^{-2}$), the sample was masked by a honeycomb grid (see insert). (1) pristine diamond, (2) area exposed only to SHI, (3) nitrogen implanted spot with NV⁻ activated by SHI, (4) NV⁻ activated by 10 keV electrons without exposure to SHI, (5) NV⁻ activated by SHI followed by exposure to 10 keV electrons, (6) nitrogen implanted area, not exposed to either SHI or electrons. Unlabeled spots and squares are from SEM imaging.

## 3. Results

We collected photoluminescence (PL) spectra for each of the six different areas in Figure 2 before (Figure 3) as well as after thermal annealing at 850° C for 1 h (Figure 4). For comparison of relative NV⁻ formation efficiencies, we subtracted the displacement damage signal from the tail of the single vacancy GR1-center line (with peak at 741 nm [16]) using data from an area that had been exposed to SHI, but was outside the nitrogen implanted spot (area 2 in Figure 2). The intensity of NV$^-$ ZPL signals from exposure to SHI was relatively low. The NV$^-$ ZPL signal also overlaps with the tail of the GR1 center signal and this led to uncertainty in the assessment of the relative NV$^-$ formation efficiency from SHI. For background correction in N-implanted regions that were masked from SHI exposures (4), we used data from pristine diamond areas (1), which we found to be spectrally equivalent to N-implanted areas prior to thermal annealing. With our given sensitivity and signal to noise, we did not observe NV$^-$ centers in pristine diamond areas or nitrogen implanted areas that had not been exposed to either SHI or electrons prior to thermal annealing.

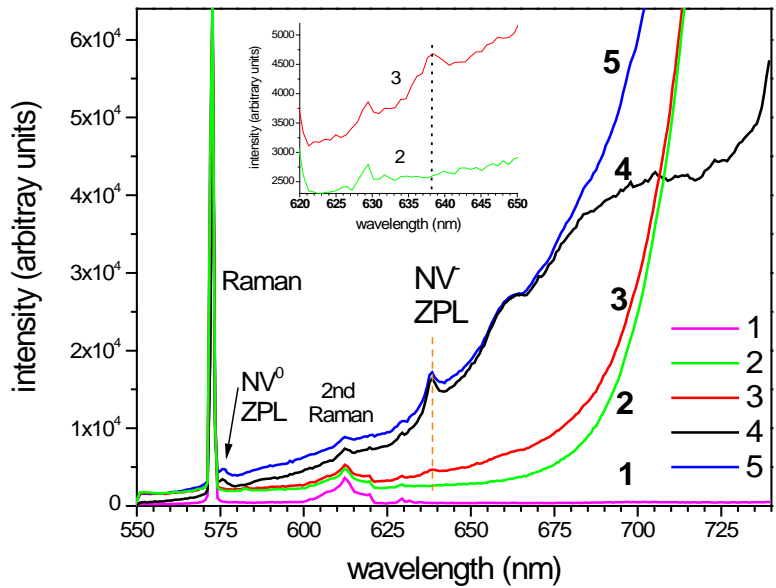

Figure 3: PL spectra of nitrogen implanted diamond (10 keV, $10^{13}$ cm$^{-2}$) with lines from NV-centers activated through 3) 1.14 GeV U-ions (5x$10^{11}$ cm$^{-2}$), 4) 10 keV electrons, 5) a combination of both (from areas as seen in Figure 1). We include spectra for 1) pristine diamond and 2) SHI induced radiation damage from outside the nitrogen implanted spot. Left to right: Primary Raman peak, NV$^0$ ZPL, secondary Raman peak, NV$^-$ ZPL, NV$^-$ side band and tail of the GR1 center signal. The spectra were recorded before

thermal annealing. The insert shows the spectral range around the NV⁻ ZPL for SHI exposure inside (3) and outside (2) the nitrogen implanted spot.

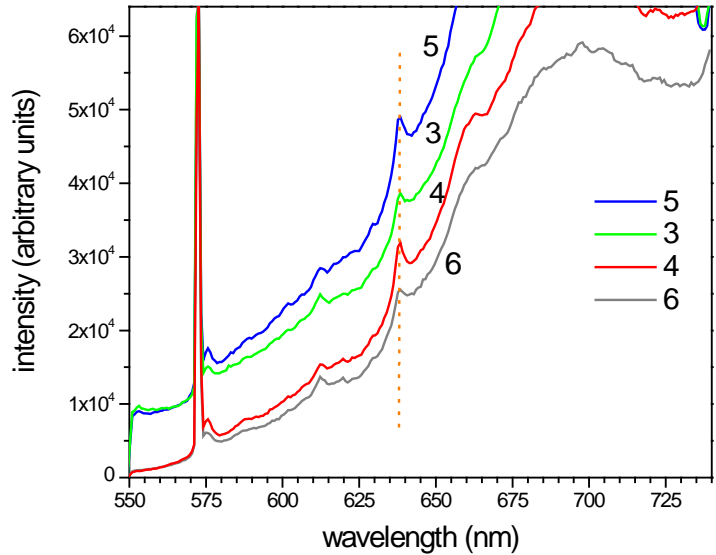

Figure 4: PL spectra for the different processing conditions presented in Figures 2 and 3 but now after thermal annealing (850°C, 1 h). Spectra showing signal from NV-centers activated through 5) 1.14 GeV U ions ($5\times10^{11}$ cm$^{-2}$) followed by electron irradiation, 3) SHI, 4) electron irradiation, 6) nitrogen implant (10 keV, $10^{13}$ cm$^{-2}$) without additional irradiations. Numbers corresponding to respective areas in Figure 2.

The passage of SHI through nitrogen implanted layers in diamond leads to the formation of NV⁻ centers (Figure 3, N + SHI) without thermal annealing. The yield of NV⁻ centers formed directly by a fluence of $5\times10^{11}$ U-ions/cm² (1.14 GeV) is about 0.1 of yields from exposure of nitrogen implanted diamonds to 10 keV energy electrons (90 C/cm²) and about 0.02 of NV⁻ yields from thermal annealing. We find the highest NV⁻ intensities when nitrogen implanted areas were first exposed to SHI and then also to low energy electrons (Figure 2, N + SHI + electrons). But this spectrum includes a contribution from the tail of the GR1 center line and after subtracting it from the NV⁻

ZPL area, we find that NV⁻ yields from electron exposures and electron + SHI are about equal (Figure 5).

After thermal annealing, the NV⁻ signal is higher in areas that had been irradiated with either SHI or keV electrons compared to areas that had been implanted with nitrogen but not irradiated further. We find that combining both irradiations and thermal annealing produces the highest NV⁻ yields (Figure 2, N + SHI + electron, and Figure 5), higher than thermal annealing alone.

To compare the PL intensities more quantitatively we subtracted the SHI induced GR1-center contribution from the NV⁻ ZPL area (635 – 642 nm) and normalized all measurements to PL signals from 10 keV N-implants ($10^{13}$ cm$^{-2}$) followed by exposure to 10 keV electrons (>90 C/cm$^2$) [17]. These data were extracted from spectra similar to those in Figure 3. We have shown earlier that the N to NV⁻ conversion efficiency from electronic activation using low energy electrons is ~25% of that from standard thermal annealing (850°C, 1 h). Thermal annealing itself has a formation efficiency for NV⁻ centers from ~10 keV nitrogen implants of only a few percent [14, 15]. The formation of NV⁻ centers by SHI (U ions, 1.14 GeV, $5\times10^{11}$ cm$^{-2}$) yields approximately 10% of the PL intensity of a 10 keV e-beam (90 C/cm$^2$). We thus estimate the absolute conversion efficiency from implanted nitrogen to NV⁻ centers, NV⁻/N, from the uranium ion irradiations to be $5\times10^{-4}$ to $10^{-3}$. In Figure 5 and 6 we compare relative NV⁻ yields before and after global annealing, from spectra as in Figure 3 and 4. The data in Figure 5 are recorded before thermal annealing and were normalized to the reference implant ($10^{13}$/cm$^2$, 10 keV N$^+$) that had also been irradiated with a saturation dose of low energy electrons [19]. The top of the error bar indicates the yields from raw data before subtraction of contributions from GR1 centers to the NV⁻ ZPL. Within the uncertainty of background subtraction, we find that exposing a nitrogen-implanted diamond first to $5\times10^{11}$ U-ions/cm² and then to low energy electrons does not yield significantly more NV⁻ centers compared to the exposure to low energy electrons alone for the sample implanted with a fluence of $10^{13}$ N/cm². Trends are similar for the 10 keV and 30 keV nitrogen implants and for the sample implanted with $10^{12}$ N/cm². However, the latter uncertainties are increased due to lower signal levels and background contributions from GR1 centers.

Data in Figure 6 were collected after thermal annealing (850°C, 1 h, in vacuum). The different processing conditions were normalized to the PL signal from thermal annealing of the nitrogen implanted diamond areas without exposure to electrons or SHI (spectrum 4 in Figure 3). Electron irradiation increases the conversion efficiency compared to standard annealing alone, up to 1.6x in the present samples. SHI and electron irradiation combined with thermal annealing show a slightly higher increase of the NV-yield of 1.7x vs. thermal annealing alone. Data from the sample exposed to $10^{11}$ Au-ions/cm² follows the same trends (not shown).

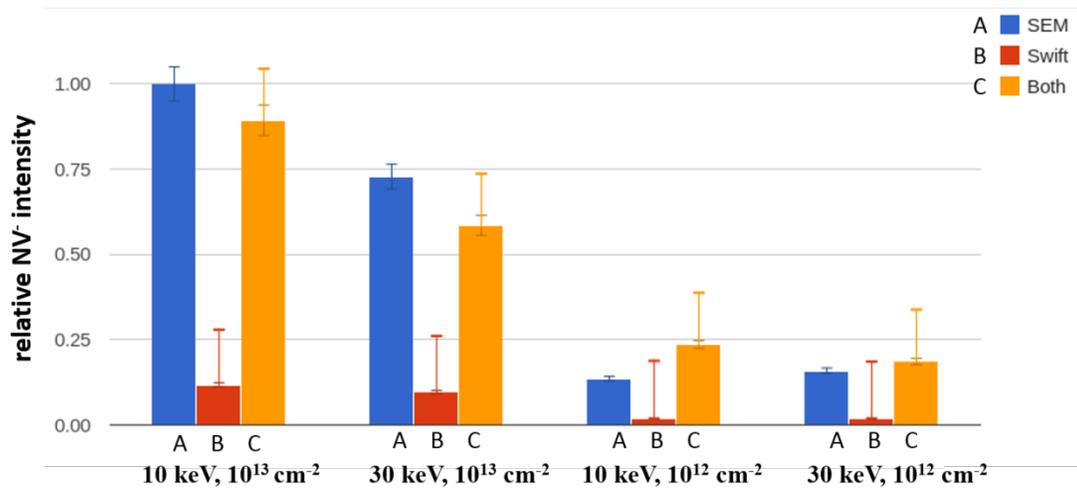

Figure 5: Relative NV⁻ PL intensities from exposure to SHI and to low energy electrons for a series of $N^+$ implantation energies and fluences without thermal annealing. For SHI alone (red) and both treatments (orange) the damage signal from SHI irradiation of pristine diamond areas was subtracted. The PL intensity before background subtraction is indicated by the top of the error bars. For the electron treatment (blue), we subtracted the background signal measured from pristine diamond. The data is normalized to NV⁻ signal from a $10^{13}/cm^2$ 10 keV $N^+$ implant followed by exposure to 10 keV electrons [19].

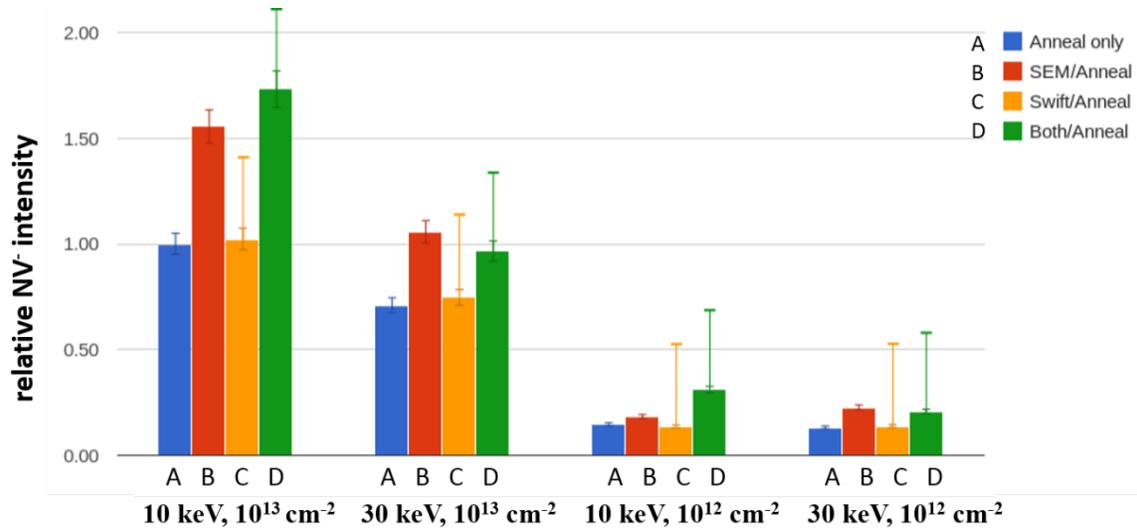

Figure 6: Relative NV$^-$ PL intensities for different fluence and energy combinations of N$^+$ implantation after thermal annealing. The data is normalized to the NV$^-$ PL intensity for a 10 keV 10$^{13}$/cm$^2$ nitrogen implant after annealing (850$^\circ$C, 1 h).

## 4. Discussion

Our earlier study showed that the formation of NV$^-$ centers by means of low-energy electron exposure does not exhibit a threshold in electron beam energy and does not depend on the e-beam current density, indicating that the formation process is not dominated by momentum transfer or target heating but rather by electronic excitations induced by the electron beam [19]. Deposition of kinetic energy of SHI along their trajectory in diamond is also dominated by electronic excitations with an electronic stopping power of ~50 keV/nm and only a small contribution from elastic collision processes (<0.1 keV/nm and ~0.25 vacancies/nm/ion [20]) in the ~130 nm deep nitrogen-implanted layer. Collisions of 1.14 GeV U-ions with target electrons can produce electrons with up to ~10 keV of kinetic energy. Subsequent electron-electron collisions lead to a cascade with a distribution of low-energy electrons. The cascade electrons thermalize within a few picoseconds, followed by radiative and non-radiative recombination or charge trapping at defect sites. The kinetics and relaxation mechanisms of the electron cascade are similar to those occurring during irradiation with keV electrons, though details of the energy, depth, and lateral distributions differ greatly. The main difference is the high excitation density of SHI with electronic energy loss rates of tens of keV/nm that can induce local thermal spike effects including localized melting and rapid re-crystallization [21-23]. We can now ask whether SHI form NV centers directly via electronic excitations similar to processes during irradiation with keV electrons, or if thermal spike effects play a role. In thermal spikes, the high electronic energy loss of projectiles can induce local melting of materials along the ion trajectory. In diamond, no permanent tracks have been observed, indicating that re-crystallization leads to at least partial lattice reconstruction following passage of SHI [21-23]. The observation of $NV^-$ centers does indicate partial lattice reconstruction of the nitrogen implanted layer in diamond following passage of SHI, while the presence of substitutional nitrogen atoms (P1 centers) is a measure of more complete damage repair. $NV^-$ can be detected by optical microscopy while the analysis of P1 centers and other non-luminescent defect complexes can be probed e.g. via spin resonance, dark state spectroscopy [1, 24] or structural probes such as transmission electron microscopy.

In order to elucidate possible mechanisms of $NV^-$ formation by SHI we now consider qualitatively electronic excitation processes from SHI and exposure to low energy electrons. For low energy electrons we found an increase of the $NV^-$ intensity with increasing electron exposure up to a saturation fluence of $\geq 10$ C/cm$^2$ [19]. If we consider the electronic stopping of 50 keV/nm for 1.14 GeV U-ions, a fluence of $5\times10^{11}$

cm$^{-2}$ and an energy for the formation of an electron-hole pair in diamond of 13 eV, then we can estimate that the SHI irradiation generates of the order of $2\times10^{17}$ electron-hole pairs/cm$^2$ in the 100 nm wide nitrogen implanted layer. For comparison to earlier measurements of the fluence dependent NV$^-$ formation from low energy electrons, the equivalent low electron fluence from SHI is then about 30 mC/cm$^2$. For low energy electrons, this fluence would corresponded to a relative NV$^-$ intensity of 500 from a 100 nm thick, nitrogen implanted layer, about 4 times lower than a relative intensity of 2000 for a saturation fluence (>10 C/cm$^2$) of low energy electrons (see Figure 2 in [19]). In the present work, we see NV$^-$ formed by SHI with a relative intensity ten times lower compared to the NV$^-$ center intensity from a high fluence of low energy electrons. Thus, within this very rough estimate, SHI appear to be 2.5x less efficient in forming NV$^-$ centers than electrons per keV of electronic energy loss in the nitrogen implanted layer.

Thermal spike effects could also play a role in NV$^-$ formation by SHI. SHI can induce thermal spikes by their high density of electronic energy deposition. Along the projectile trajectory, a hot cylindrical zone is created with a heating and cooling time of a few ps. Simulations predict that temperatures well above the (pressure dependent) stability limit of NV$^-$ centers of about 1200 to 2000$^{\circ}$C [25] can be reached within the cylinder core of ~5 nm radius [21-23]. This hot core is surrounded by progressively cooler regions. Within an area of about 200 nm$^2$ the temperature can still be above 600$^{\circ}$C and induce vacancy mobility for a brief period. It seems plausible that SHI can stimulate NV$^-$ formation in a region surrounding the ion trajectory that is heated, but not to a temperature above the NV stability limit. Here, NV$^-$ centers can form by the two-step process of nitrogen incorporation into the lattice followed by capture of a mobile vacancy, and by a combination of electronic excitation and thermal activation for reconstruction of defect complexes that contain interstitial nitrogen and vacancies. Our present data does not allow us to draw any quantitative conclusions on the relative importance of these processes. At the given fluence for U ions of $5\times10^{11}$ cm$^{-2}$ and an estimated heated area per ion of ~200 nm$^2$, all nitrogen ions (implanted with a with fluence of $10^{13}$ cm$^{-2}$) experience both a local thermal transient together with electronic excitations. Studies of NV$^-$ formation as a function of electronic stopping power and fluence can elucidate the interplay of electronic excitation and thermal spike effects.

Following SHI irradiation and keV-electron exposures, we performed a standard thermal annealing process (Figure 6). After annealing, the NV$^-$ yields are higher in areas that were exposed to electrons or first exposed to SHI and then to electrons, compared to thermal annealing alone [19]. SHI irradiations form NV's directly, but with relatively low efficiency. Apparently, SHI irradiation also modifies the structure of the nitrogen implanted layer such that consecutive exposure to keV electrons produces slightly more NV centers than electron exposure alone, without SHI pre-treatment. For the 10 keV nitrogen implant with fluence of $10^{13}$ cm$^{-2}$, we find that the combination of SHI, followed by electron beam and finally thermal annealing produces the highest NV$^-$ yields, more than thermal annealing alone, and more than electron exposure plus thermal annealing. For the 30 keV nitrogen implant, NV$^-$ formation by 10 keV electrons is less efficient, probably because electrons loose more energy before reaching the peak of the implanted nitrogen distribution. Trends are similar for nitrogen implants with $10^{12}$ cm$^{-2}$, but signal levels were lower.

Both low energy electrons and SHI can also affect defect complexes of NV

centers with hydrogen [18], but we did not characterize the hydrogen content of the diamonds used here. This is likely a small effect in the present study, as our starting material was similar to the diamonds used in Ref. 17, where hydrogen related NV$^-$ PL quenching effects were observed following specific hydrogen plasma treatments.

**5. Outlook and conclusions**

One challenge in the realization of a quantum computer architecture with NV$^-$ center qubits is the formation of arrays with ~10 nm period so that the NV$^-$-NV$^-$ magnetic dipolar coupling strength is sufficient to enable gate control and error correction much faster than the de-phasing time, $T_2$, or the formation of spin chains consisting of both P1 and NV$^-$ centers [1, 2]. The following simple steps enable the formation of three dimensional NV$^-$ - N assemblies with SHI. First, electronic grade diamond is implanted with a selected depth profile of nitrogen ions, e. g. using multiple implantation energies. In a second step, the diamond is irradiated with SHI and NV$^-$ centers are formed along the ion trajectories. Assemblies of NV$^-$ can be tailored with a distribution of NV$^-$ spacings resulting from the local nitrogen concentrations and N to NV$^-$ conversion efficiencies.

Local NV$^-$ formation without global thermal annealing also enables a try-and-repeat approach to NV$^-$ array formation. After a single nitrogen ion is implanted and registered in a selected location [5, 9], an NV$^-$ center can be formed by low energy electron irradiation, by targeted SHI irradiation [26], or by another form of local excitation [27, 28]. The location can be repeatedly probed for the presence of an NV$^-$ center by PL or CL. Once an NV$^-$ is detected, the process is repeated at the next location or more nitrogen and vacancies can be introduced by local ion implantation, followed by local excitation until an NV-center is observed. In this way, arrays of single NV centers can be formed. Coherence times of locally formed $NV^-$ have to be quantified to evaluate their potential uses.

$NV^-$ yields from exposure of nitrogen implanted diamonds to SHI and keV electrons followed by a standard thermal annealing step are ~1.7x higher than yields from thermal annealing alone. But absolute conversion yields, $NV^-/N$, are still low, of order a few percent. Our results show that the simple model of $NV^-$ formation in nitrogen implanted diamond in a two-step process of P1 center formation followed by vacancy trapping at substitutional nitrogen sites is incomplete. SHI can form $NV^-$ centers along tracks like "pearls on a string" over distances of tens of µm's. Local $NV^-$ formation by SHI or low-energy electrons enables try-and-repeat approaches for the deterministic formation of NV center arrays in which nitrogen is first implanted and then locally activated. Further, our finding of NV formation by the passage of SHI in pre-implanted and pre-damaged diamond shows that $NV^-$ centers are probes of (partial) lattice damage recovery. Tracking of $NV^-$ centers and other color centers may be useful in studies of lattice damage and recovery dynamics in the interplay of elastic and inelastic energy loss process during irradiations [29-31].

**Acknowledgements**

This work was performed in part at the Molecular Foundry and the National Center for Electron Microscopy at Lawrence Berkeley National Laboratory and was supported by the Office of Science, Office of Basic Energy Sciences, Scientific User

**Corresponding author contact information:** T_Schenkel@lbl.gov, 510-486-6674